\documentstyle[12pt]{article}
  \textwidth                 
                             16.4cm
  \oddsidemargin              
                              2.5cm
  \advance\oddsidemargin 
                             by -1in
  \evensidemargin             
                              0.0cm
  \advance\evensidemargin
                             by -1in
  \marginparwidth 
                              1.9cm
  \marginparsep
                              0.4cm
  \marginparpush 
                              0.4cm
  \topmargin 
                             -3.0cm
  \advance\topmargin 
                             by -0.0in
  \textheight 
                             25.0cm
\makeindex

  \newcommand\beq{\begin{equation}}
  \newcommand\noi{\noindent}
  \newcommand\eeq{\end{equation}}
  \newcommand\beqn{\begin{eqnarray}}
  \newcommand\eeqn{\end{eqnarray}}

  \def\Pom{{\bf I\!P}}
  
  \def\lsim{\mathrel{\rlap{\lower4pt\hbox{\hskip1pt$\sim$}}
      \raise1pt\hbox{$<$}}}         %less    than or approx. symbol
  \def\gsim{\mathrel{\rlap{\lower4pt\hbox{\hskip1pt$\sim$}}
      \raise1pt\hbox{$>$}}}         %greater than or approx. symbol
\begin{document}

\vspace*{10mm}
\begin{center} 
%%%%%%%%%%%%%%%%%%%%%%%%%%%%%%%%%%%%%%%%%%%%%%%% 
{\LARGE \bf Is It Possible to Study the Wave Function \\
\vspace*{2mm}
of $2S$ Vector Mesons at HERA ?}
%%%%%%%%%%%%%%%%%%%%%%%%%%%%%%%%%%%%%%%%%%%%%%%% 

\vspace*{10.0mm}

{{\Large Jan Nemchik}}

\end{center}
%\maketitle
  \bigskip
\begin{center}
  \noi
 {\sl 
  Institute of Experimental Physics SAS, \\
  Watsonova 47,  \\
  04353 Kosice,  \\
  Slovakia}
  
\end{center}
  \bigskip
  \vspace*{3.0cm}

%***************  
\begin{abstract}
%***************
We present a short review of anomalous properties in diffractive photo- and
electroproduction of radially excited $V'(2S)$ vector mesons.
Using the color dipole gBFKL phenomenology we analyze
anomalous $Q^{2}$ and energy dependence of the 
production cross section, $V'(2S)/V(1S)$ production
ratio, the diffraction slope and anomalous
$t$ behaviour of the differential cross section $d\sigma/dt$.
The origin of these anomalies is based on the interplay of the nodal
structure of $V'(2S)$ radial wave function with the energy and dipole
size dependence of the dipole cross section and the diffraction slope.
We analyze how a different pattern of anomalous behaviour of $V'(2S)$
production leads to a different position of the node in the wave function 
and discuss how that node position can be extracted from the data at HERA.   
%*************  
\end{abstract}
%*************
% ----------------------------------------------------------------------
  
\newpage
%\doublespace
%
%
%*********************** SECTION 1 ***************************
%  
%
%  
%*********************  
\section{Introduction}
%*********************

The main goal of this paper is to present a short
review of possible anomalies, which can be observed
in diffractive photo-
and electroproduction
of radially excited $V'(2S)$ vector mesons
%
%
% ===============================
\beq
\gamma^{*}p \rightarrow V'(2S)p ~~~~~~~~V'(2S) = \rho', 
\Phi', \omega', \Psi', \Upsilon',...\, .
\eeq
% ===============================
%
%
Diffractive electroproduction of ground state $V(1S)$
vector mesons at high c.m.s. energy $W = \sqrt{s}$
is intensively discussed during the last decade
and is very convenient for study of the pomeron exchange
\cite{DL,KZ91,Ryskin,KNNZ93,KNNZ94,NNZscan,Brodsky,Forshaw,GLM}.
The standard approach to the pQCD is based on the BFKL equation
\cite{Kuraev,Balitsky,Lipatov} formulated
in the scaling approximation of the infinite gluon correlation
radius $R_{c}\rightarrow\infty$ (massless gluons) and
of the fixed running coupling $\alpha_{S}=const$.
Later, however, a novel $s$-channel approach 
to the $LLs$ BFKL equation (running gBFKL approach)
has been developed \cite{NZ94,NZZ94} in terms of the color
dipole cross section
$\sigma(\xi,r)$ ($r$ is the transverse size of
the color dipole, $\xi = log ({{W^{2}+Q^{2}} \over {m_{V}^{2}+Q^{2}}})$
is the rapidity variable) and incorporates consistently
the asymptotic freedom (AF) (i.e. the running QCD coupling $\alpha_{S}(r)$)
and the finite propagation radius $R_{c}$ of perturbative gluons.

As a consequence of the gBFKL phenomenology for
diffractive production of light
\cite{NNZscan,NNPZ97} and heavy
\cite{NNPZZ98} vector mesons is so-called 
{\it scanning phenomenon} \cite{NNN92,KNNZ93,KNNZ94,NNZscan};
the
$V(1S)$ vector meson production amplitude probes the color dipole cross
section at the dipole size $r\sim r_{S}$,
where the scanning radius $r_{S}$ can be expressed through the scale
parameter $A$
%
% ---------------------------------------------------------------------
\beq
r_{S} \approx {A \over \sqrt{m_{V}^{2}+Q^{2}}}\, ,
\label{eq:1}
\eeq
% ---------------------------------------------------------------------
%
where $Q^{2}$ is the photon virtuality,
$m_{V}$ is the vector meson mass, $A\approx 6$
and slightly rises with $Q^{2}$.
Consequently,
changing $Q^{2}$ and the mass of the produced
vector meson one can
probe the dipole cross section $\sigma(\xi,r)$
in a very broad range of the dipole
sizes $r$.
This fact allows to study the transition
from large nonperturbative dipole size $r_{S}\gg R_{c}$
to the perturbative region of very short $r_{S}\ll R_{c}$.
Futhermore, 
the scanning 
phenomenon give a possibility to study one important 
consequence of the color dipole gBFKL   
dynamics -
the steeper
subasymptotic energy dependence of the dipole cross section
at smaller dipole sizes ${\bf r}$.

Diffractive electroproduction
of radially excited $V'(2S)$ vector mesons supplies us
with an additional
information on the dipole cross section.
The presence of the node in
$V'(2S)$ radial
wave function leads to
a strong cancellation
of dipole size contributions to the production amplitude
from the region above and below the node position
$r_{n}$ (the node effect
\cite{KZ91,NNN92,KNNZ93,NNZanom,NNPZ97,NNPZZ98}).
For this reason the amplitudes for
electroproduction of the $V(1S)$ and $V'(2S)$ vector mesons probe
$\sigma(\xi,r)$ in a different way.

The node effect as the dynamical mechanism has a lot of
very interesting consequences in production of $V'(2S)$
vector mesons and can be tested at HERA.
Therefore, we present a short review of different aspects
and manifestations of the node effect. 
Firstly, we show the onset of a strong node effect in
electroproduction of $V'(2S)$
light vector mesons \cite{NNPZ97}, which leads to
a very spectacular pattern
of anomalous $Q^{2}$ and energy dependence of production
cross section. More heavy is the vector meson
much weaker is the node effect.
However, for electroproduction of $V'(2S)$ heavy vector mesons 
much weaker node effect 
still leads to a slightly different $Q^{2}$- and
energy dependence of production cross section
for $\Psi'$ vs. $J/\Psi$
and to a nonmonotonic $Q^{2}$- dependence of
the diffraction slope at small
$Q^{2}\lsim 5$\,GeV$^{2}$ for $\Psi'$ production
\cite{NNPZZ98}.
Then we discuss 
another manifestation of the node effect
experimentally confirmed at fixed target and HERA
experiments in $J/\Psi$ and $\Psi'$ photoproduction;
a strong
suppression of diffractive production of $V'(2S)$ vs. $V(1S)$ mesons.
The stronger is the node effect the smaller is  the $V'(2S)/V(1S)$ 
production ratio.
The node effect 
in conjunction with the emerging gBFKL
phenomenology of the diffraction slope
\cite{NZZslope,NZZspectrum,NNPZZ98}
also leads to a counterintuitive inequality
$B(\gamma^{*}\rightarrow \Psi') \lsim B(\gamma^{*}\rightarrow J/\Psi)$
\cite{NNPZZ98}, which can be also tested at HERA.
However, we show that above counterintuitive inequality $B(2S)<B(1S)$
is not always valid for $V'(2S)$ light vector meson production
\cite{Nbd00}.
Finally, we analyze the node effect and its manifestation
resulting in a very spectacular pattern of anomalous $t$ dependence
of the differential cross section \cite{Ntd00}.
In all the cases the main emphasize will be related to the production of
$V'(2S)$ light vector mesons where the node effect is
expected to be very strong. Also we find a 
correspondence between a specific pattern of anomalous behaviours
and the position of the node in
$V'(2S)$ radial wave function.

The paper is organized as follows. In Sect.~2 we present
a very short review of the color dipole phenomenology of
diffractive photo- and electroproduction of vector mesons
In Sect.~3 we analyze the anomalies in production
cross section for $V'(2S)$ vector mesons.
Sect.~4 is devoted to anomalous pattern of $Q^{2}$
and energy dependence of the diffraction slope for
$V'(2S)$ production.
In Sect.~5 we study anomalous $t$ behaviour of the 
differential cross section
$d\sigma(\gamma^{*}\rightarrow V'(2S))/dt$ at different $Q^2$
and energies.
In all the cases we
discuss how the position of the node
in $V'(2S)$ radial wave function can be extracted from the
data. The summary and conclusions are presented in Sect.~6.
  
%
%
%*********************** SECTION 2 ***************************
%  
%
%  
%*******************************************************  
\section{Color dipole phenomenology
         for vector meson production. A short review.}
%*******************************************************

The light-cone representation introduced in \cite{ks70}
represents
very popular and powerful tool for study of the dynamics of
vector meson diffractive photo- and electroproduction. 
The central point of this approach is that
in the mixed $({\bf{r}},z)$ representation
the high energy vector meson can be treated
as a system of color dipole described by
the distribution
of the transverse separation ${\bf{r}}$ of the quark and
antiquark given by the $q\bar{q}$ wave function,
$\Psi({\bf{r}},z)$, where $z$ is
the fraction of meson's light-cone momentum 
carried by a quark.
In this approach the
imaginary part of the production
amplitude for the real (virtual) photoproduction
of vector mesons
with the momentum transfer ${\bf{q}}$ can be represented in the
factorized form
% ---------------------------------------------------------------------
\beq
{\rm Im}{\cal M}(\gamma^{*}\rightarrow V,\xi,Q^{2},{\bf{q}})=
\langle V |\sigma(\xi,r,z,{\bf{q}})|\gamma^{*}\rangle= 
\int\limits_{0}^{1} dz\int d^{2}{\bf{r}}\sigma(\xi,r,z,{\bf{q}})
\Psi_{V}^{*}({\bf{r}},z)\Psi_{\gamma^{*}}({\bf{r}},z)\,
\label{eq:2}
\eeq
% ---------------------------------------------------------------------
whose normalization is
$
\left.{d\sigma/ dt}\right|_{t=0}={|{\cal M}|^{2}/ 16\pi}.
$
In Eq.~(\ref{eq:3}),
$\Psi_{\gamma^{*}}({\bf{r}},z)$ and
$\Psi_{V}({\bf{r}},z)$ represent the
probability amplitudes
to find the color dipole of size $r$
in the photon and quarkonium (vector meson), respectively.
The color dipole distribution in (virtual) photons was
derived in \cite{NZ91,NZ94}.
$\sigma(\xi,r,z,{\bf{q}})$ in Eq~(\ref{eq:3})
is the dipole scattering matrix for $q\bar{q}-N$ interaction.
At ${\bf{q}}=0$ it represents
the color dipole cross section, which quantifies
the interaction of the relativistic
color dipole of the dipole size ${\bf{r}}$ with the
target nucleon.
The dipole cross section $\sigma(\xi,r)$ is flavor
independent and represents the universal
function of $r$ which describes
various diffractive processes in unified form.
Energy dependence of the dipole cross section
reflexes an importance of the higher Fock states $q\bar{q}g...$
at high c.m.s. energy $W$.
In the leading-log ${1\over x}$ approximation the
effect of higher Fock states can be
reabsorbed into the energy dependence
of $\sigma(\xi,r)$, which satisfies
the gBFKL equation
\cite{NZ94,NZZ94} for the energy evolution.

At small ${\bf{q}}$ considered in this paper,
one can safely neglect
the $z$-dependence of $\sigma(\xi,r,z,{\bf{q}})$
for light and heavy vector meson production
and set $z=\frac{1}{2}$.
This follows also from the analysis within double gluon
exchange approximation
\cite{NZ91} leading to a slow $z$ dependence of
the dipole cross section.

The detailed discussion about the space-time
pattern of diffractive electroproduction of vector mesons
is presented in \cite{NNPZ97,NNPZZ98}.
The energy dependence of the dipole cross section is quantified
in terms of the dimensionless
rapidity $\xi=\log{1\over x_{eff}}$, $x_{eff}$ is
the effective value of the Bjorken variable
%
%
% ===============================================================
\beq
x_{eff} =
\frac {Q^{2}+m_{V}^{2}}{Q^{2}+W^{2}} \approx  
 \frac{m_{V}^{2}+Q^{2}}{2\nu m_{p}}\, ,
\label{eq:3}
\eeq
% ===============================================================
%
%
where $m_{p}$ and $m_{V}$ is the proton mass and mass of
vector meson, respectively.
Hereafter, we will write the energy dependence of the dipole
cross section in both variables,
either in $\xi$ or in $x_{eff}$ whenever convenient.

The production amplitudes for the
transversely (T) and the longitudinally (L) polarized vector mesons
with the small momentum transfer $\bf{q}$
can be written in more explicit form \cite{NNZscan,NNPZZ98}
% =================================================================
\beqn
{\rm Im}{\cal M}_{T}(x_{eff},Q^{2},{\bf{q}})=
{N_{c}C_{V}\sqrt{4\pi\alpha_{em}} \over (2\pi)^{2}}
\cdot~~~~~~~~~~~~~~~~~~~~~~~~~~~~~~~~~ 
\nonumber \\
\cdot \int d^{2}{\bf{r}} \sigma(x_{eff},r,{\bf{q}})
\int_{0}^{1}{dz \over z(1-z)}\left\{
m_{q}^{2}
K_{0}(\varepsilon r)
\phi(r,z)-
[z^{2}+(1-z)^{2}]\varepsilon K_{1}(\varepsilon r)\partial_{r}
\phi(r,z)\right\}
\label{eq:4}
\eeqn
%=================================================================
\beqn
{\rm Im}{\cal M}_{L}(x_{eff},Q^{2},{\bf{q}})=
{N_{c}C_{V}\sqrt{4\pi\alpha_{em}} \over (2\pi)^{2}}
{2\sqrt{Q^{2}} \over m_{V}}
\cdot~~~~~~~~~~~~~~~~~~~~~~~~~~~~~~~~~ 
 \nonumber \\
\cdot \int d^{2}{\bf{r}} \sigma(x_{eff},r,{\bf{q}})
\int_{0}^{1}dz \left\{
[m_{q}^{2}+z(1-z)m_{V}^{2}]
K_{0}(\varepsilon r)
\phi(r,z)-
\partial_{r}^{2}
\phi(r,z)\right\} 
\label{eq:5}
\eeqn
%==================================================================
where
%==================================================================
\beq
\varepsilon^{2} = m_{q}^{2}+z(1-z)Q^{2}\,,
\label{eq:6} 
\eeq
%==================================================================
$\alpha_{em}$ is the fine structure
constant, $N_{c}=3$ is the number of colors,
$C_{V}={1\over \sqrt{2}},\,{1\over 3\sqrt{2}},\,{1\over 3},\,
{2\over 3},\,{1\over 3}~~$ for
$\rho^{0},\,\omega^{0},\,\phi^{0},\, J/\Psi, \Upsilon$ production,
respectively and
$K_{0,1}(x)$ are the modified Bessel functions.
The discussion and parameterization
of the light-cone radial wave function $\phi(r,z)$
of the $q\bar{q}$ Fock state of the vector meson
is given in \cite{NNPZ97}.

Following the scanning phenomenon (see Eq.~(\ref{eq:1})) 
one needs rather high values of $Q^2\gsim 70$\, GeV$^{2}$ to reach the pure
perturbative region $r\lsim R_{c}$.
Consequently,
due to a very slow onset of the pure perturbative region
one can easily anticipate
a contribution to the production amplitude
coming
from the semiperturbative and nonperturbative $r\gsim R_{c}$.
Following the simplest assumption about an additive property
of the perturbative and nonperturbative mechanism of interaction
we can represent the contribution of the bare pomeron exchange
to $\sigma(\xi,r,{\bf{q}})$ as a sum
of the perturbative and nonperturbative component  
%
%===============================================================
\footnote{
additive property of such a decomposition of the dipole
cross section has been already discussed in \cite{NNPZ97,NNPZZ98}}
%===============================================================
%
%
% ==============================================================
\beq
\sigma(\xi,r,{\bf{q}}) =
\sigma_{pt}(\xi,r,{\bf{q}})+\sigma_{npt}(\xi,r,{\bf{q}})\,,
\label{eq:7}
\eeq  
% ==============================================================
%
%
with the parameterization of both components at small ${\bf{q}}$
% ==============================================================
\beq
\sigma_{pt,npt}(\xi,r,{\bf{q}})=\sigma_{pt,npt}(\xi,r,{\bf{q}}=0)
\exp\Bigl(-\frac{1}{2}
B_{pt,npt}(\xi,r){\bf{q^{2}}}\Bigr)\,.
\label{eq:8}
\eeq
% ==============================================================
Here $\sigma_{pt,npt}(\xi,r,{\bf{q}}=0)
= \sigma_{pt,npt}(\xi,r)$ represent the contribution
of the perturbative and nonperturbative mechanism to the
$q\bar{q}$-nucleon interaction cross section,
respectively, $B_{pt}(\xi,r)$ and $B_{npt}(\xi,r)$
are the corresponding dipole diffraction slopes.

The model predictions include also a 
small real part of production amplitudes taken
in the form \cite{GribMig}
%
%
%------------------------------------------------
\beq
{\rm Re}{\cal M}(\xi,r) =\frac{\pi}{2}\cdot\frac{\partial}
{\partial\xi} {\rm Im}{\cal M}(\xi,r)\,.
\label{eq:9}
\eeq

The formalism for calculation of $\sigma_{pt}(\xi,r)$
in the leading-log $s$ approximation was developed
in \cite{NZ91,NZ94,NZZ94}.
The contribution $\sigma_{npt}(\xi,r)$
representing the soft nonperturbative component
of the pomeron is a simple Regge pole with
the intercept $\Delta_{npt}=0$.
The particular form together with   
assumption of
the energy independent
$\sigma_{npt}(\xi=\xi_{0},r)=\sigma_{npt}(r)$
($\xi_{0}$ corresponds to boundary condition for the gBFKL
evolution, $\xi_{0}=log(1/x_{0})$, $x_{0} = 0.03$)
allows us to successfully describe the
proton structure function at very small $Q^{2}$ \cite{NZHera},
the real photoabsorption \cite{NNZscan} and
diffractive real and virtual photoproduction of light
\cite{NNPZ97} and heavy \cite{NNPZZ98} vector mesons.
Besides, the reasonable form of $\sigma_{npt}(r)$
was confirmed in the process of the first determination of the dipole
cross section from the data on vector meson
electroproduction \cite{NNPZdipole} what is shown in Fig.~1. The energy
and dipole size dependence of so-extracted $\sigma(\xi,r)$
is in a good agreement with the dipole cross section obtained
from the gBFKL dynamics \cite{NNZscan,NZHera}.
The  nonperturbative component of the pomeron
exchange plays a dominant
role at low NMC energies
in the production of the light vector mesons.
However, the perturbative component of the pomeron becomes
more important with the rising energy also in the nonperturbative
region of dipole sizes.

%
%
%------------------------- FIG.1. ----------------------------
%
%  
\begin{figure}[tbh]
  \includegraphics{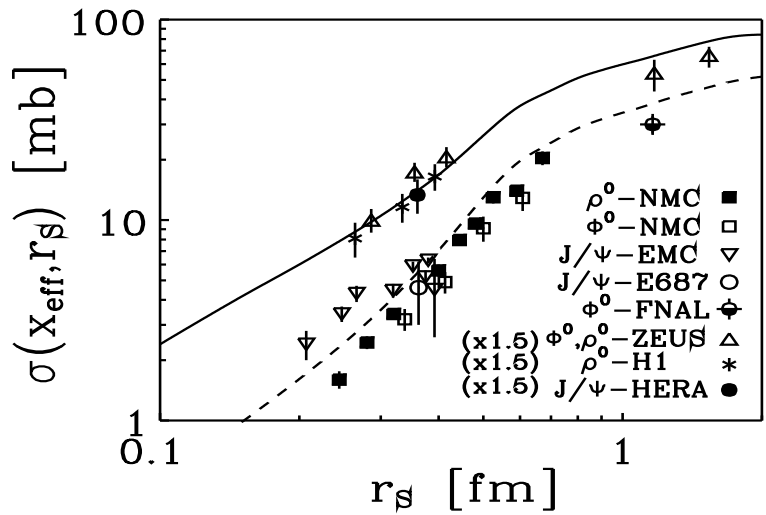}
  \begin{center}
  \vspace{6.2cm}
  \parbox{13cm}
  {\caption[Delta]
  {
The dipole size dependence of the dipole cross section
extracted from the data on photoproduction
and electroproduction of vector mesons.
The data are described in \cite{NNZscan}.
The dashed and solid curve show the dipole cross section
of the model \cite{NZHera,NNZscan} evaluated for
the c.m.s. energy $W=15$ and $70$\, GeV respectively.
The data points at HERA energies and the corresponding solid curve
are multiplied by the factor 1.5.
 }
%-------------------
       \label{event1}}
%-------------------
  \end{center}
  \end{figure}
%------------------------------------------------------------- 
%
%

The generalization of the color dipole factorization
formula (\ref{eq:2}) to the diffraction slope of the
reaction $\gamma^{*}p\rightarrow Vp$ reads \cite{NNPZZ98}
%
%
%------------------------------------------------
\beq
B(\gamma^{*}\rightarrow V,\xi,Q^{2})
{\rm Im} {\cal M}(\gamma^{*}\rightarrow V,\xi,Q^{2},{\bf{q}}=0)=
\int\limits_{0}^{1} dz\int d^{2}{\bf{r}}\sigma(\xi,r)~B(\xi,r)
\Psi_{V}^{*}(r,z)\Psi_{\gamma^{*}}(r,z)\,.
\label{eq:10}
\eeq
%------------------------------------------------
%
%

The diffraction cone in the color dipole gBFKL approach
was studied in detail in \cite{NNPZZ98}. 
Therefore, here we present only the salient
feature of the color diffraction slope $B(\xi,r)$ emphasizing  
the presence of the geometrical contribution from beam
dipole - $r^{2}/8$
and the contribution from the target proton size - $R_{N}^{2}/3$:
%
%
%------------------------------------------------
\beq
B(\xi,r)=
\frac{1}{8}r^{2}+\frac{1}{3}R_{N}^{2}+
2\alpha_{\Pom}'(\xi-\xi_{0}) + {\cal O}(R_{c}^{2})\, ,
\label{eq:11}
\eeq
%------------------------------------------------
%
%
where $R_{N}$ is the radius of the proton.
The term $2\alpha_{\Pom}'(\xi-\xi_{0})$
describe the familiar Regge growth of $B(\xi,r)$ for
the quark-quark scattering.
The geometrical contribution to the diffraction  
slope from the target proton size ${1\over 3}R_{N}^{2}$
persists for all the dipole sizes
$r\gsim R_{c}$ and $r\lsim R_{c}$. The last term in (\ref{eq:11})
is also associated with the proton size and is negligibly small.
The diffractive scattering of large color dipole has
been also studied in the paper \cite{NNPZZ98}.
Here we assume the conventional Regge rise of the diffraction
slope for the soft pomeron \cite{NNPZZ98}
%
%
%------------------------------------------------ 
\beq
B_{npt}(\xi,r)=\Delta B_{d}(r)+\Delta B_{N}+
2\alpha_{npt}^{'}(\xi-\xi_{0})\,,
\label{eq:12}
\eeq
%------------------------------------------------
%
%
where $\Delta B_{d}(r)$ and $\Delta B_{N}$ stand for the contribution
from the beam dipole and target nucleon size.
As a guidance the 
data on the pion-nucleon scattering
\cite{Schiz} were used, which suggest $\alpha'_{npt}=0.15$\,GeV$^{-2}$.
In (\ref{eq:12}) the proton size contribution
is
%
%
%------------------------------------------------
\beq
\Delta B_{N}={1\over 3}R_{N}^{2}\, ,
\label{eq:13}
\eeq
%------------------------------------------------
%
%   
and
the beam dipole contribution has been proposed
to have a form \cite{NNPZZ98}
%
%
%------------------------------------------------
\beq
\Delta B_{d}(r) = {r^{2} \over 8}\cdot
{r^{2}+aR_{N}^{2} \over 3r^{2}+aR_{N}^{2}}\,,
\label{eq:14}
\eeq
%------------------------------------------------
%
%   
where $a$ is a phenomenological parameter, $a\sim 1$.
We take $\Delta B_{N}=4.8\,{\rm GeV}^{-2}$.
Then the pion-nucleon diffraction slope is reproduced with
reasonable value of the parameter $a$ in the formula (\ref{eq:14}):
$a=0.9$ for $\alpha'_{npt}=0.15$\,GeV$^{-2}$.
    
Energy dependence of the
gBFKL diffraction slope $B(\xi,r)$ (see Eq.~(\ref{eq:11})
and \cite{NZZslope}) can be evaluated
through the energy
dependent effective Regge slope $\alpha_{eff}'(\xi,r)$
%
%
% ===================================================================
\beq
B_{pt}(\xi,r) \approx \frac{1}{3}<R_{N}^{2}> + \frac{1}{8}r^{2}
+ 2\alpha_{eff}'(\xi,r)(\xi-\xi_{0}).
\label{eq:15}
\eeq
% ====================================================================
%
%
The effective Regge slope $\alpha_{eff}'(\xi,r)$
varies
with energy differently
at different dipole size
\cite{NZZslope}.
At fixed scanning radius and/or $Q^{2}+m_{V}^{2}$,
it decreases with energy.
At fixed rapidity $\xi$
and/or $x_{eff}$ (\ref{eq:3}),
$\alpha_{eff}'(\xi,r)$
rises with $r\lsim 1.5$\,fm.
At fixed energy it is a flat function
of the scanning radius.
At asymptotically large $\xi$ ($W$),
$\alpha_{eff}'(\xi,r)\rightarrow \alpha_{\Pom}'=0.072$\,GeV$^{-2}$.
At lower and HERA energies the subasymptotic
$\alpha_{eff}'(\xi,r)\sim (0.15-0.20)$\,GeV$^{-2}$ and is very
close to $\alpha_{soft}'$ known from the Regge phenomenology
of soft scattering.
It means that the gBKFL dynamics predicts a substantial rise
with the energy and dipole size of the diffraction slope $B(\xi,r)$
in accordance with
the energy and dipole size dependence of the effective
Regge slope $\alpha_{eff}'(\xi,r)$ and due to a presence of the
geometrical components $\propto r^{2}$ in (\ref{eq:15}) and
$\Delta B_{d}(r)\propto r^{1.7}$
in (\ref{eq:12}) (see also (\ref{eq:14}))
% ==========================================
\footnote{Dipole size behaviour of
$\Delta B_{d}(r)$ (\ref{eq:14}) representing the geometrical
contribution to the
dipole diffraction slope $B_{npt}(\xi,r)$ (\ref{eq:12}) for
diffractive scattering of large color dipole has the standard
$r^{2}$- dependence at small $r^{2}\ll aR_{N}^{2}$
and large $r^{2}\gg aR_{N}^{2}$ values of dipole size,
respectively.
In the intermediate region $r^{2}\sim aR_{N}^{2}$,
which corresponds to production of $V(1S)$ and $V'(2S)$ light
vector mesons, the dipole size dependence of $\Delta B_{d}(r)$
can be parameterized by the power function $r^{\alpha}$
with $\alpha\sim 1.7$.}.
% ==========================================
The overall dipole diffraction slope contains contributions
from both $B_{npt}(\xi,r)$ and $B_{pt}(\xi,r)$ and
corresponding geometrical component has $r^{\alpha}$-
behaviour with $1.7< \alpha\lsim 2.0$.
Therefore, for discussions on the qualitative level
in the subsequent sections we
assume (with a reasonable accuracy) 
an approximate $r^{2}$- dependence of the
geometrical component contribution
to the dipole diffraction slope.
The first direct evaluation of the dipole diffraction slope
from the data on photo- and electroproduction of vector mesons
is presented in \cite{Nbsys00} and is depicted in Fig.~2.

%
%
%------------------------- FIG.2. ----------------------------
%
%  
\begin{figure}[tbh]
  \includegraphics{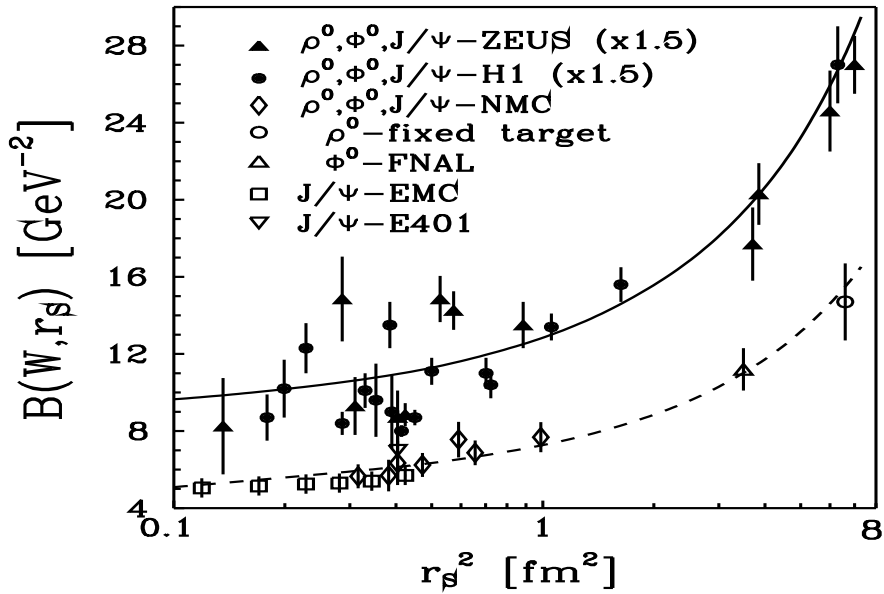}
  \begin{center}
  \vspace{7.5cm}
  \parbox{13cm}
  {\caption[Delta]
  {
The dipole size dependence of the dipole cross section
extracted from the data on photoproduction
and electroproduction of vector mesons.
The data are described in \cite{Nbsys00}.
The dashed and solid curve show the dipole diffraction slope
of the model \cite{NZZslope,NNPZZ98} evaluated for
the c.m.s. energy $W=15$ and $70$\, GeV respectively.
The data points at HERA energies and the corresponding solid curve
are multiplied by the factor 1.5.
 }
%-------------------
       \label{event2}}
%-------------------
  \end{center}
  \end{figure}
%------------------------------------------------------------- 
%
%

%%%%%%%%%%%%%%%%%%%%%%%%%%%%%%%%%%%%%%%%%%%%%%%%%%%%%%%%%%%%%%%
%
%
%
%*********************** SECTION 3 ***************************
%  
%
%  
%****************************************  
\section{Anomalous cross section in electroproduction of $2S$
          radially excited vector mesons}
%****************************************

The matrix element for $V'(2S)$ diffractive production  
contains the contributions
from the region of dipole sizes
above and below the node position $r_{n}$.
As soon as the exact node effect encounters 
the $Q^{2}$- and energy dependent cancellations
from the soft (large size) and hard (small size)
contributions to the $V'(2S)$ production amplitude
become important.
The strong $Q^{2}$ dependence of the
node effect is connected with the $Q^{2}$ behaviour
of the scanning radius $r_{S}$ (see (\ref{eq:1})).
The energy dependence of the cancellations comes from
a different energy dependence of the dipole cross section
$\sigma(\xi,r)$ at different dipole sizes $r$.  

We would like to emphasize from the very beginning
that the predictive power is weak and the predictions
are strongly model dependent
in the region of $Q^{2}$ and energy
when the node effect becomes exact.
Presenting and discussing in the subsequent sections
the model predictions for $V'(2S)$ vector mesons
we do not insist on the precise pattern of an anomalous
behaviour. We present the model
calculations only 
as an illustration of possible anomalies, which can be
tested at HERA
%
%================================================================
\footnote{
Manifestations of the node
effect in electroproduction on nuclei were discussed earlier, see
\cite{NNZanom} and \cite{BZNFphi}}.
%================================================================
%
We will concentrate mainly on the production of $V'(2S)$ 
light vector mesons
because of a strong node effect and the fact that
the new data obtained at HERA will be analyzed soon.
In the nonrelativistic limit of heavy quarkonia, the
node effect will not depend on the polarization of the virtual photon
and of the produced vector meson. Not so for light vector mesons
\cite{NNPZ97}.
The wave functions of $(T)$ and $(L)$ polarized (virtual)
photon are different.
Different regions of $z$ contribute to the
${\cal M}_{T}$ and ${\cal M}_{L}$.
Different scanning radii
for production of $(T)$ and $(L)$ polarized vector mesons
and different energy dependence of $\sigma(\xi,r)$ at
these scanning radii
lead to a different $Q^{2}$ and energy dependence of the
node effect in production of $(T)$ and $(L)$ polarized
$V'(2S)$ vector mesons.

There are two possible scenarios for the node effect:
the undercompensation
and the overcompensation regime \cite{NNZanom}.
In the undercompensation scenario,
the $V'(2S)$ production amplitude
$\langle V'(2S)|\sigma(\xi,r)|\gamma^*\rangle$
is dominated by the positive valued contribution coming from small
dipole sizes $r\lsim r_{n}$
and the $V(1S)$ and $V'(2S)$ photoproduction
amplitudes have the same sign.
This scenario corresponds namely to the production
of $V'(2S)$ heavy vector mesons ($\Psi'(2S)$, $\Upsilon'(2S),...)$.
In the overcompensation scenario,
the $V'(2S)$ production amplitude
is dominated by the negative valued contribution coming from large
dipole sizes $r\gsim r_{n}$,
and the $V(1S)$ and $V'(2S)$ photoproduction
amplitudes have the opposite sign.
This scenario can correspond to the production of $V'(2S)$
light vector mesons, $\rho'(2S)$, $\omega'(2S)$ and $\phi'(2S)$
%
%================================================================
\footnote{discussion on the experimental determination of the
relative sign of the $V'(2S)$ and $V(1S)$ production amplitudes
using the so-called S\" oding-Pumplin effect \cite{SP1,SP2}
has been already presented in \cite{NNPZ97}}. 
%================================================================ 
%

Let us start with (T) polarization. 
In the undercompensation scenario
\cite{NNZanom} a decrease of
of the scanning radius
with $Q^{2}$ leads to
a rapid decrease of the negative contribution coming from
large $r\gsim r_{n}$ and to a rapid
rise of the $V'(2S)/V(1S)$ production ratio with $Q^{2}$.
The stronger the suppression of the real photoproduction of
the $V'(2S)$ state, the steeper the $Q^{2}$ dependence
of the $V'(2S)/V(1S)$ production ratio expected at small $Q^{2}$.
In Fig.~3 we predict 
the $\rho'(2S)/\rho^{0}$ and $\phi'(2S)/\phi^{0}$ (T) polarized
production ratios using the wave functions from \cite{NNPZ97}.
They rise by more than one order of magnitude
in the range $Q^{2}\lsim 0.5$\,GeV$^{2}$.
At larger $Q^{2}\gsim 1$\,GeV$^{2}$, when the production   
amplitudes are dominated by dipole size $r\ll r_{n}$
the $V'(2S)$ and $V(1S)$
production cross sections become comparable.
\cite{NNZanom,NNZscan}.

%
%
%------------------------- FIG.3. ----------------------------
%
%  
\begin{figure}[tbh]
  \includegraphics{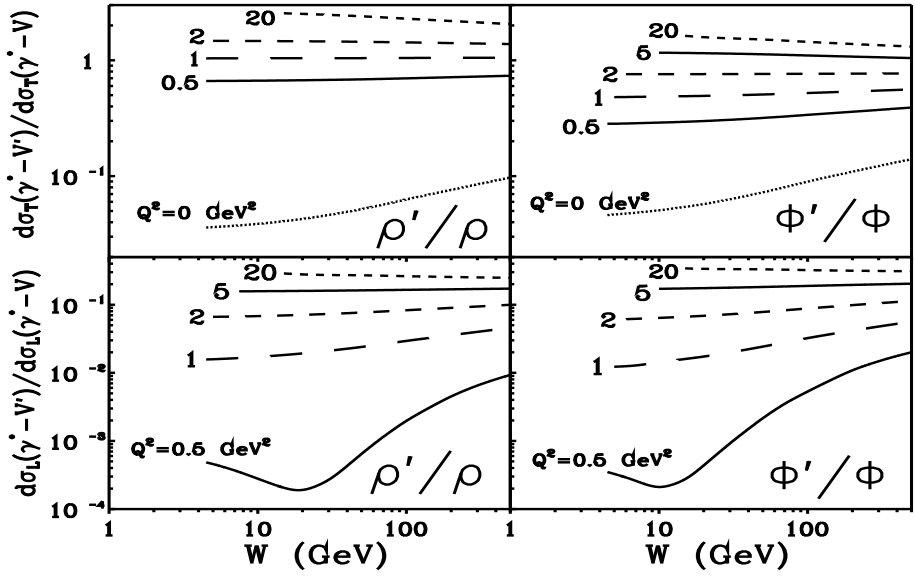}
  \begin{center}
  \vspace{6.9cm}
  \parbox{13cm}
  {\caption[Delta]
  {
The color dipole model
predictions for the $Q^2$ and $W$ dependence of the ratios
$\sigma(\gamma^{*}\rightarrow \rho'(2S))/
\sigma(\gamma^{*}\rightarrow \rho^{0})$ and
$\sigma(\gamma^{*}\rightarrow \phi'(2S))/
\sigma(\gamma^{*}\rightarrow \phi^{0})$
for the (T) and (L)
polarization of the vector mesons.
 }
%-------------------
       \label{event3}}
%-------------------
  \end{center}
  \end{figure}
%------------------------------------------------------------- 
%
%

Using the wave functions from \cite{NNPZ97} we predict
the overcompensation scenario at $Q^{2}=0$
for $(L)$ polarized $\rho'(2S)$ and $\phi'(2S)$ mesons.
Consequently,
the decrease with $Q^{2}$ of the scanning
radius $r_{S}$  leads to the {\sl exact} cancellation
of the small and large distance contributions at some value
$Q_{c}^{2}\sim 0.5$\,GeV$^{2}$ for both the $\rho'_{L}(2S)$ and
$\phi'_{L}(2S)$ production.
The value of $Q_{c}^{2}$ is
slightly different for the imaginary and the real part of
$V'(2S)$ production amplitude. 
We can not insist on the precise value of
$Q_{c}^{2}$ which is subject to the soft-hard cancellations,
our emphasis is on the likely scenario with the exact node
effect at a finite $Q_{c}^{2}$.

At larger $Q^{2}$ and/or smaller  
scanning radius one enters the above described
undercompensation scenario.      
For both $(T)$ and $(L)$
polarized photons, $V'(2S)/V(1S)$ production ratios rise steeply with $Q^{2}$
on the scale $Q^{2}\sim 0.5$\,GeV$^{2}$.
At large $Q^{2}$ where the production of
$(L)$ polarized mesons dominates,  the
$\rho'(2S)/\rho^{0}$ and $\phi'(2S)/\phi^{0}$ cross section ratios
level off at $\sim 0.3-0.4$ (see Fig.~3).
This large-$Q^{2}$
limiting value of the
production cross section ratios
depends on the ratio of $V'(2S)$ and $V(1S)$ wave functions
at the origin, which in potential models is subject to
the detailed form of the confining potential
\cite{Potential}.

%
%
%------------------------- FIG.4. ----------------------------
%
%  
\begin{figure}[tbh]
  \includegraphics{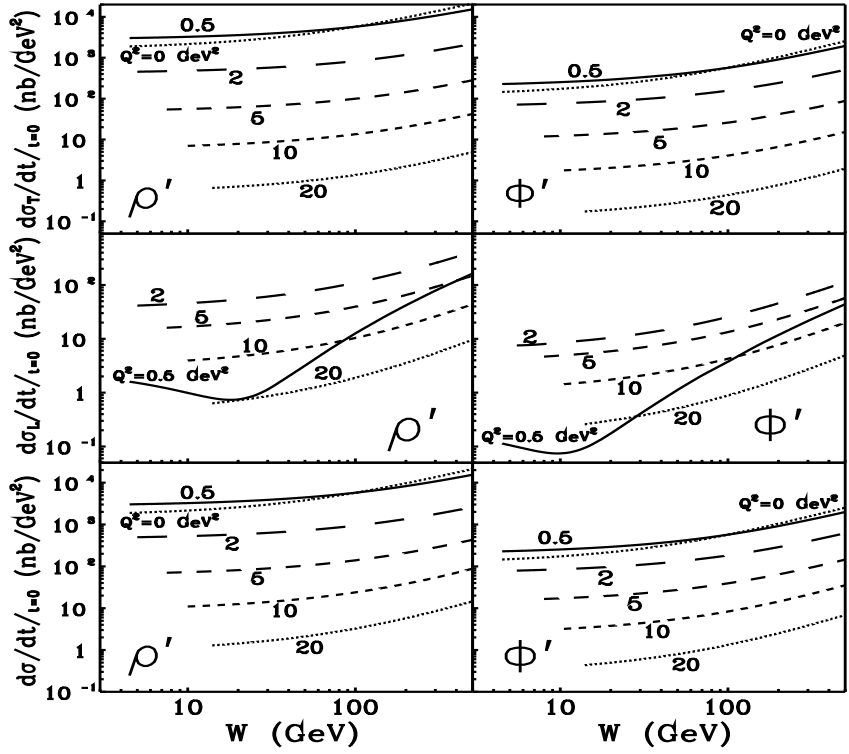}
  \begin{center}
  \vspace{9.3cm}
  \parbox{13cm}
  {\caption[Delta]
  {
The color dipole model
predictions of the forward differential cross sections   
$d\sigma_{L,T}(\gamma^* \rightarrow V')/dt|_{t=0}$ for
transversely(T)
(top boxes) and longitudinally (L)
(middle boxes) polarized radially excited
vector mesons $\rho'(2S)$ and
$\phi'(2S)$ and for the polarization-unseparated
$d\sigma(\gamma^* \rightarrow V')/dt|_{t=0}=
d\sigma_{T}(\gamma^* \rightarrow V')/dt|_{t=0}+\epsilon
d\sigma_{L}(\gamma^* \rightarrow V')/dt|_{t=0}$ for
$\epsilon = 1$ (bottom boxes)
as a function of the c.m.s. energy $W$
at different values of $Q^2$.
}
%-------------------
       \label{event4}}
%-------------------
  \end{center}
  \end{figure}
%------------------------------------------------------------- 

The energy dependence of the $\rho'(2S)$ and $\phi'(2S)$
real photoproduction is shown in Fig.~4 and has again a
specific pattern. In the color dipole gBFKL dynamics,
the negative contribution
to the $V'(2S)$ production amplitude coming
from 
$r\gsim r_{n}$,
has a slower growth with energy than the positive
contribution coming
from $r\lsim r_{n}$. Consequently,
in the undercompensation regime (which corresponds to $(T)$
polarization)
the destructive interference
of these two contributions becomes weaker at higher energy and
we predict a growth with energy of the
$V'_{T}(2S)/V_{T}(1S)$ production cross section ratios  (see Fig.~3)
and $(T)$ polarized forward production cross sections (see Fig.~4 -
top boxes).

For $(L)$ polarized $V_{L}'(2S)$ 
we have a chance of
studying the $Q^{2}$ and energy dependence in the overcompensation
scenario. 
At moderate energy and
$Q^{2}$ very close to $Q_{c}^{2}$ but
still $\lsim Q_{c}^{2}$ the negative
contribution from $r\gsim r_{n}$ still takes over in the $V_{L}'(2S)$
production amplitude. With increasing energy, the
positive contribution to the production amplitude
rises faster and gradually takes over. At some
intermediate energy there is an exact cancellation
of the two contributions to the production
amplitude and 
$V_{L}'(2S)$ production cross section 
exhibits a minimum. The value of the minimum is finite
because cancellations in the real and imaginary part of the  
production amplitude are different.
Using the model wave functions from \cite {NNPZ97}
we find such a nonmonotonic energy dependence of the
$\rho_{L}'(2S)$ and $\phi_{L}'(2S)$
production at $Q^{2}\approx 0.5$\,GeV$^{2}$,
which is shown in Fig.~3 (bottom boxes) and Fig.~4 (middle boxes).
At higher $Q^{2}>Q_{c}^{2}$ one encounters the above described 
undercompensation
scenario and
the energy dependence of $V_{L}'(2S)/V_{L}(1S)$
production ratios and of $V'_{L}(2S)$ production
cross sections becomes very weak.

%
%
%*********************** SECTION 4 ***************************
%  
%
%  
%*********************************************
\section{Anomalous diffraction slope for production of $2S$
         radially excited vector mesons}
%*********************************************

For a
better understanding of anomalous properties
of the $V'(2S)$ diffraction slope,
the generalized factorization formula (\ref{eq:10})
can be rewritten as the ratio of two matrix elements
%
%
% ---------------------------------------------------------------------
\beqn
B(\gamma^{*}\rightarrow V(V'),\xi,Q^{2},{\bf q}=0) =
\frac{
\langle V(V') |\sigma(\xi,r)B(\xi,r)|\gamma^{*}\rangle
}{
\langle V(V') |\sigma(\xi,r)|\gamma^{*}\rangle} =
\nonumber \\
\frac{
\int\limits_{0}^{1} dz\int d^{2}{\bf{r}}\sigma(\xi,r)
B(\xi,r)\Psi_{V(V')}^{*}({\bf{r}},z)\Psi_{\gamma^{*}}({\bf{r}},z)
}{
\int\limits_{0}^{1} dz\int d^{2}{\bf{r}}\sigma(\xi,r)
\Psi_{V(V')}^{*}({\bf{r}},z)\Psi_{\gamma^{*}}({\bf{r}},z)} = \frac{{\cal
N}}
{{\cal D}}
\, ,
\label{eq:16}
\eeqn
% ---------------------------------------------------------------------
%
%
where ${\cal N}$ and ${\cal D}$ denotes the numerator and
denominator, respectively.

Anomalous properties of the diffraction slope comes namely
from (\ref{eq:16}).
The denominator ${\cal D}$ represents the well-known production
amplitude.
As it was mentioned
the $V(1S)$ production amplitude
is dominated by contribution from
dipole size $r\sim r_{S}$ (\ref{eq:1}).
However,
because of an approximate $\propto r^{2}$
behaviour of the slope parameter
the integrand of the matrix element in the numerator ${\cal N}$
of Eq.~(\ref{eq:22})
is dominated by the dipole size
$r = r_{B}\sim 5/3r_{S}$.

Let ${\cal M}_{+}$ and ${\cal M}_{-}$ be the moduli
of positive and negative valued contributions to the $V'(2S)$
production amplitude from the region
of dipole sizes $r < r_{n}$ and $r > r_{n}$, and let
$B_{+}$ and $B_{-}$ be the diffraction slopes for the
corresponding contributions. Because of an approximate
$\sim r^{2}$ dependence of the diffraction slope (see discussion
in Sect.~2) we have a strong inequality
%
%
%------------------------------------------------
\beq
B_{+} < B_{-}
\label{eq:17}
\eeq
%------------------------------------------------
%
%
The overall
$V'(2S)$ production amplitude is 
${\cal M}(2S) = {\cal M}_{+}-{\cal M}_{-}$ and the 
corresponding overall diffraction slope for $V'(2S)$
production reads
%
%
%------------------------------------------------
\beqn
B(2S) = \frac{B_{+}{\cal M}_{+} - B_{-}{\cal M}_{-}}
{ {\cal M}_{+}-{\cal M}_{-}}\nonumber \\
=
B_{+} - (B_{-} - B_{+})\frac{{\cal M}_{-}}
{{\cal M}_{+}-{\cal M}_{-}} \,  ,
\label{eq:18}
\eeqn
%------------------------------------------------
%
%
which can be rewritten in a more convenient
form for the following discussion
%
%
%------------------------------------------------   
\beq
B(2S) - B(1S) =
- (B_{-} - B_{+})\frac{{\cal M}_{-}}
{{\cal M}_{+}-{\cal M}_{-}} \,  ,
\label{eq:19}
\eeq
%------------------------------------------------
%
%
where $B(1S)\approx B_{+}$ for production of $V(1S)$
vector mesons.
For the diffractive production of $V'(2S)$ heavy vector
mesons the production amplitude
${\cal M}(2S)$ is
positive valued (undercompensation scenario)
at $Q^{2}=0$ and consequently we predict   
from (\ref{eq:19}) a counterintuitive
inequality  $B(\Psi'(2S)) < B(J/\Psi(1S))$ \cite{NNPZZ98}
although the r.m.s. radius of $\Psi'$ is much larger than $R_{J/\Psi}$.
However, we will manifest below that this is not     
always true in production of $V'(2S)$ light vector mesons.

%
%
%------------------------- FIG.5. ----------------------------
%
%  
\begin{figure}[tbh]
  \includegraphics{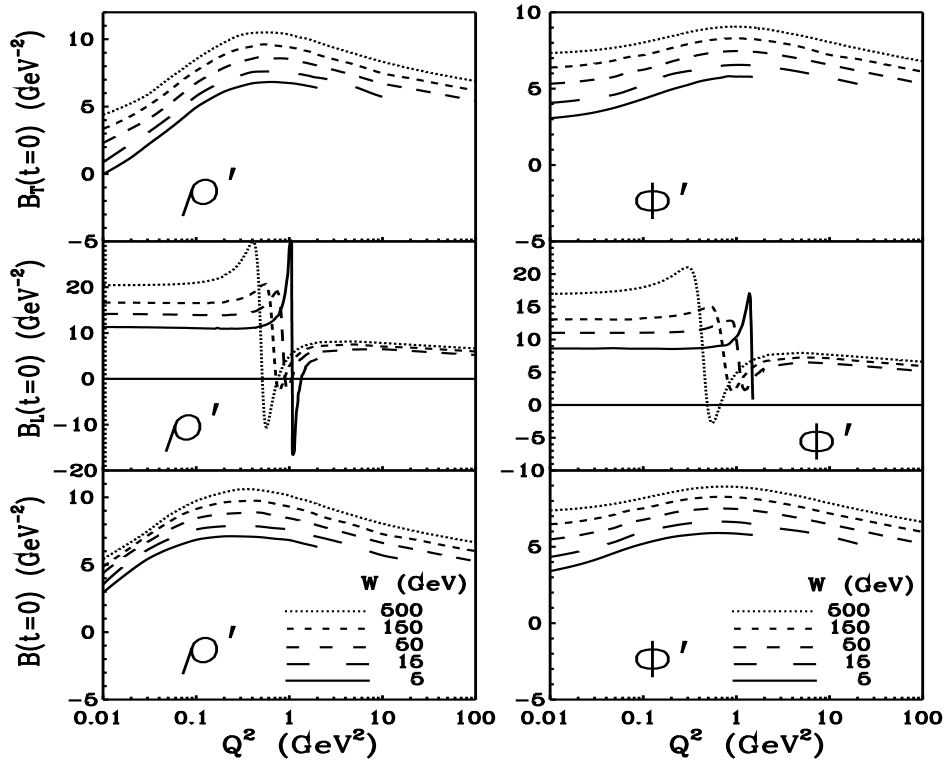}
  \begin{center}
  \vspace{9.30cm}
  \parbox{13cm}
  {\caption[Delta]
  {
~- The color dipole model
predictions for the $Q^{2}$ dependence of the diffraction slope $B(t=0)$
for production of
transversely (T)
(top boxes), longitudinally (L)
(middle boxes) polarized
and polarization-unseparated
(T) + $\epsilon$(L) (bottom boxes)
$\rho'(2S)$ and $\phi'(2S)$
for $\epsilon = 1$
at different values of the c.m.s. energy $W$.
     }
%-------------------
       \label{event5}}
%-------------------
  \end{center}
  \end{figure}
%------------------------------------------------------------- 
%
\vspace*{3mm}

{\bf Undercompensation scenario}  

The undercompensation scenario
(${\cal D} > 0$ in (\ref{eq:16}))
corresponds to the
production of $(T)$ polarized $\rho'(2S)$ and $\phi'(2S)$
at $Q^{2}=0$ using the
wave functions from Ref.~\cite{NNPZ97}.
Because of $r_{B}> r_{S}$,
there are two possibilities concerning the sign of  
${\cal N}$ in Eq.~(\ref{eq:16}). \\
i.)
     ${\cal N} < 0$; ${\cal N}$ and ${\cal D}$ have
     the opposite sign. Consequently, the diffraction
     slope in the photoproduction limit is negative
     valued. This pattern corresponds to diffractive
     photoproduction of $\rho_{T}'(2S)$ at small energy (see Figs.~5
     and 6 - top boxes). \\
ii.)
     ${\cal N} > 0$; ${\cal N}$ and ${\cal D}$ have
     the same sign. Consequently, the diffraction
     slope in the photoproduction limit is positive
     valued. This pattern corresponds to diffractive
     photoproduction of $\phi_{T}'(2S)$ (see Figs.~5 and 6 -
     top boxes) and the node effect is weaker
     in comparison with $\rho_{T}'(2S)$ photoproduction.\\  
In both cases we predict from Eq.~(\ref{eq:19}) the 
counterintuitive inequalities
$B(\rho_{T}'(2S)) < B(\rho_{T}(1S))$ and
$B(\phi_{T}'(2S)) < B(\phi_{T}(1S))$,
which are analogical
to that for charmonium diffractive photoproduction \cite{NNPZZ98}.
     
A decrease of the scanning radius with $Q^{2}$ leads to a very
rapid decrease of the negative valued contribution to the
diffraction slope
coming from $r\gsim r_{n}$ and consequently
leads to a steep rise of $B(V_{T}'(2S))$ with $Q^{2}$.
The higher is $Q^{2}$ the weaker is the node effect and
the smaller is the difference
$|B(2S) - B(1S)|$.
At still larger $Q^{2}$ and
at fixed energy $W$
the slope parameter $B(V_{T}'(2S))$
exhibits a broad maximum
at some value of $Q_{T}^{2}\in
(0.5-2.0)$\,GeV$^{2}$. 
At very large $Q^{2}\gg m_{V}^{2}$ when the node effect becomes
negligible, $B(2S)\sim B(1S)$ and
$B(V_{T}'(2S))$ decreases monotonously with $Q^{2}$
following the $Q^{2}$ dependence
of $B(V_{L,T}(1S))$.
The above described pattern of nonmonotonic $Q^{2}$ dependence
of the diffraction slope is
depicted in Fig.~5 (bottom boxes) for both the
$\rho_{T}'(2S)$ and $\phi_{T}'(2S)$ production.

\vspace*{3mm}

{\bf Overcompensation scenario}

The overcompensation scenario
(${\cal D} < 0$ in (\ref{eq:16})),
corresponds to the
production of $(L)$ polarized $\rho'(2S)$ and $\phi'(2S)$
at $Q^{2}=0$ using the
wave functions from Ref.~\cite{NNPZ97}.
Because of $r_{B} > r_{S}$,
${\cal N} < 0$ and has the same sign as ${\cal D}$.
Consequently, the diffraction slope in the
photoproduction limit is positive valued as it can be 
expected also from the
undercompensation
regime described above. The sign of the diffraction slope $B(2S)$
at $Q^{2}=0$ can not distinguish between the overcompensation
and undercompensation scenarios.
However, because of ${\cal M}(2S) < 0$ the difference
$B(2S) - B(1S)$ is positive valued (see Eq.~(\ref{eq:19}).
As the result we predict the expected inequalities
$B(\rho_{L}(1S)) < B(\rho_{L}'(2S))$ and
$B(\phi_{L}(1S)) < B(\phi_{L}'(2S))$, what is a new result
in comparison with the color dipole predictions for heavy
vector mesons presented in the paper \cite{NNPZZ98}.

With the decrease of the scanning radius with $Q^{2}$
there is a rapid decrease of the negative contributions
to ${\cal N}$ and ${\cal D}$   
coming from $r\gsim r_{n}$.
For some $Q^{2}\sim Q'^{2}_{L}\in (0.5-1.5)$\,GeV$^{2}$
one encounters the exact node effect firstly for the denominator 
${\cal D}$ and $B(V_{L}'(2S))$ has a
peak for both the $\rho_{L}'(2S)$ and
$\phi_{L}'(2S)$ production.
The value of $B(V_{L}'(2S))$ corresponding to
this exact node effect will be finite due to a different node effect
for the real and imaginary part of the production amplitude.
The onset of the node effect causes also the rapid
continuous transition of $B(V_{L}'(2S))$ from positive to  
negative values
when the matrix element ${\cal D}$ passes from the
overcompensation to undercompensation regime.
Consequently, for $Q^{2} > Q'^{2}_{L}$,
${\cal D} > 0$, ${\cal N}$ is kept to be negative valued
and $B(V_{L}'(2S))$ starts to
rise from its minimal negative value (see Fig.~5 - middle
boxes).
At still larger $Q^{2}$ the    
following pattern of the $Q^{2}$ behaviour of
$B(V_{L}'(2S))$ is analogical to that for
$Q^{2}$ dependence of $B(V_{T}'(2S))$.
For the production of polarization unseparated $V'(2S)$,
the anomalous properties of $B(V_{L}'(2S))$ are essentially
invisible and the corresponding slope parameter
$B(V'(2S))$ has an analogical $Q^{2}$ dependence as $B(V_{T}'(2S))$
and is shown in Fig.~5 (bottom boxes).
Here we can not insist on the precise value of 
$Q'^{2}_{L}$ which is again a subject of the soft-hard cancellations.
We would like only to emphasize that the exact node effect
for $B(V_{L}'(2S))$ is at a finite $Q'^{2}_{L}$.

To conclude 
the 
$Q^{2}$ dependence of $B(2S)$,
the undercompensation scenario
is characterized by a broad maximum
at $Q^{2}\sim Q^{2}_{T}$ and can
be tested experimentally at HERA measuring the virtual
photoproduction of the $\rho'(2S)$ and $\phi'(2S)$ at
$Q^{2}\in (0-10)$\, GeV$^{2}$ and at different values of energy.
However, the overcompensation scenario
is characterized by a sharp peak followed by a very rapid transition of
the diffraction slope from positive to negative values  
at $Q^{2}\sim Q'^{2}_{L}$ and then by a broad maximum at 
$Q^{2}\sim Q^{2}_{L}$ and
can be also investigated at HERA separating $(L)$ polarized
$\rho_{L}'(2S)$ and $\phi_{L}'(2S)$ at moderate
$Q^{2}\in (0.1-5.0)$\,GeV$^{2}$.

%
%------------------------- FIG.6. ----------------------------
%
%  
\begin{figure}[tbh]
  \includegraphics{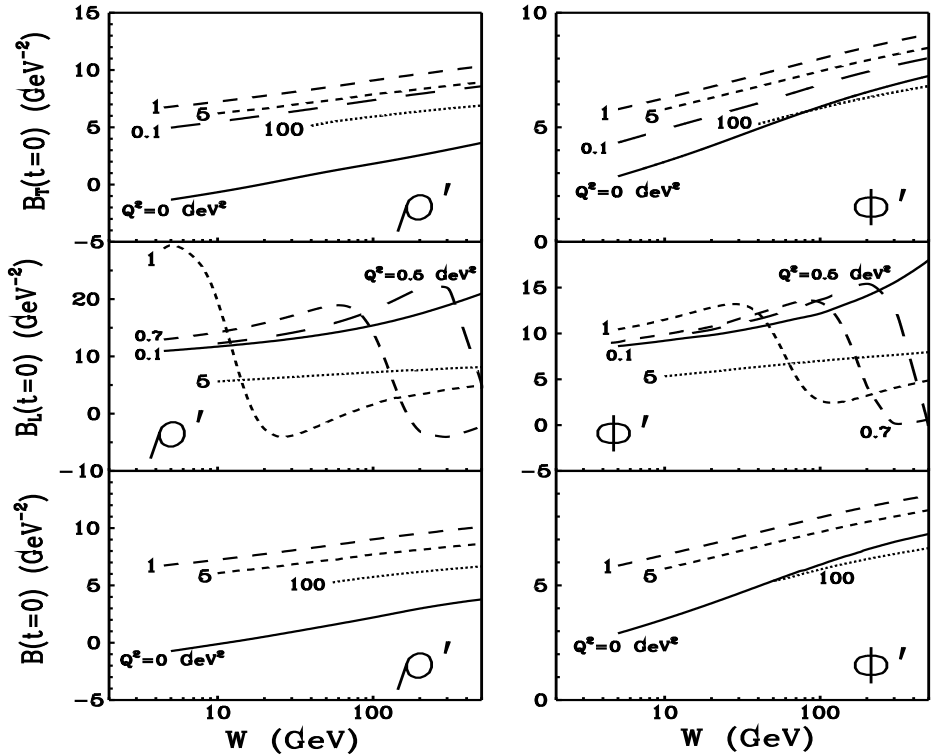}
  \begin{center}
  \vspace{9.5cm}
  \parbox{13cm}
  {\caption[Delta]
  {
~- The color dipole model
predictions for the $W$ dependence of the diffraction slope $B(t=0)$
for production of
transversely (T)
(top boxes), longitudinally (L)
(middle boxes) polarized
and polarization-unseparated 
(T) + $\epsilon$(L) (bottom boxes)
$\rho'(2S)$ and $\phi'(2S)$
for $\epsilon = 1$
at different values of $Q^2$.
     }
%-------------------
       \label{event6}}
%-------------------
  \end{center}
  \end{figure}
%------------------------------------------------------------- 

The energy dependence of $B(V'(2S))$
at different $Q^{2}$ is shown
in Fig.~6 and has its own peculiarities.
Let us start with $B(V_{T}'(2S))$ at $Q^{2}=0$
when the production amplitude is in the undercompensation regime,
${\cal D} > 0$.
Fig.~6 demonstrates (top boxes) steeper rise with energy
of the diffraction slope at lower $Q^{2}$ and confirms
the expectation coming from the gBFKL dynamics that 
the energy dependence of $B_{T}'(2S)$
is given mainly by the effective Regge slope $\alpha'$
(see Eqn.~(\ref{eq:12}) and (\ref{eq:15}),
which decreases with $Q^{2}$.

The successful separation of $(L)$ polarized
$V_{L}'(2S)$ mesons at HERA offers an unique possibility
to study an anomalous energy dependence of
the diffraction slope in the overcompensation scenario.
At $Q^{2}=0$ 
both the ${\cal N}$ and ${\cal D}$ of Eq.~(\ref{eq:16})
are negative valued.
At moderate energy and $Q^{2}$ close but smaller than
$Q'^{2}_{L}$, the negative valued contribution
coming from $r\gsim r_{n}$ still takes over in ${\cal D}$
(${\cal N} < 0$ as well due to $r_{B}> r_{S}$).
Because of a steeper rise with energy of
the positive contribution to the $V'(2S)$ production amplitude
coming from $r\lsim r_{n}$
than the negative contribution coming from
$r\gsim r_{n}$, we find an exact cancellation
of these two contributions to ${\cal D}$ and a maximum
of the diffraction slope $B(V_{L}'(2S))$ at some intermediate energy
followed by its rapid continuous transition
from the positive to negative values,
when ${\cal D}$ passes
from the overcompensation to the undercompensation regime.
Different node effect for the real and imaginary part of
the production amplitude provides such a continuous transition.
At larger energies
${\cal D} > 0$ (undercompensation regime) and consequently  
$B(V_{L}'(2S))$ is negative valued and starts to rise 
from its minimal negative value.
Such a situation is depicted in Fig.~6 (middle boxes), where we
predict with the wave functions from Ref.~\cite{NNPZ97}
such a nonmonotonic energy behaviour
of $B(V_{L}'(2S))$ for both $\rho'(2S)$ 
and $\phi'(2S)$ production at $Q^{2}\lsim Q'^{2}_{L}\in (0.5-1.5)$\, 
GeV$^{2}$.
The position of the maximum $W_{t}$
and the transition from the positive to negative
values of $B(V_{L}'(2S))$ depends on $Q^{2}$
and can be measured at HERA.
At higher $Q^{2}$ and smaller scanning radii,
the further pattern of the energy behaviour for
$B(V_{L}'(2S)$ is analogical to that
for $B(V_{T}'(2S))$.

If $(T)$ and $(L)$
polarized $V_{T}'(2S)$ and $V_{L}'(2S)$ mesons will be 
separated experimentally there is a chance for
experimental determination
of a concrete scenario in $(T)$ and $(L)$
polarized $V'(2S)$ production amplitude.
The simplest test can be realized in the photoproduction limit
($Q^{2}=0$) for a broad energy range.
If the data will report the
counterintuitive inequality $B(2S) < B(1S)$
($B(2S)$ can be also negative valued) then
$V'(2S)$ production amplitude is in the undercompensation
regime (positive valued). In the opposite case when
the expected inequality $B(2S) > B(1S)$ will be obtained from the data then
$V'(2S)$ production amplitude is in the overcompensation
regime (negative valued).
For the production of $(L)$ polarized vector mesons
the values of $Q^{2}$
should be high enough to have the data with a reasonable
statistics however, must not be very large in order
to have a considerably strong node effect. We propose the range
of $Q^{2}\in (0.5-5.0)$\,GeV$^{2}$ for exploratory study
of the overcompensation scenario at HERA.

%***************************************
\section{Anomalous $t$ dependence of the differential cross section
         for production of $2S$ radially excited vector mesons}
%***************************************  
  
We discuss now the possible
peculiarities in $t$ dependent differential cross
section $d\sigma/dt$ for $V'(2S)$ production.
We would like to emphasize again
that we do not insist on
the precise form of the $t$ dependence of $d\sigma/dt$,
the main emphasis is on the likely pattern
of the $t$ dependence coming from the node effect.
The differential cross section is calculated 
using the expressions (\ref{eq:4}) and (\ref{eq:5}) for
$(T)$ and $(L)$ production amplitudes in conjunction
with Eqs.~(\ref{eq:7}), (\ref{eq:8}), (\ref{eq:12}) and (\ref{eq:15}).
Because of an approximate $\propto r^{2}$ behaviour of
the geometrical contribution to the diffraction slope (see 
discussion in Sect.~2),
the large size negative contribution to the production
amplitude
from the region $r>r_{n}$
corresponds to larger value of the diffraction slope than the
small size contribution from the region $r<r_{n}$.
It means that
the negative contribution to the $V'(2S)$ production amplitude
has a steeper $t$ dependence than the positive contribution.
Let $t$-dependent production amplitude be 
%
%
%-------------------------------------------------------------------------
\beq
 {\cal M}(t) = c_{+}\exp(-\frac{1}{2}B_{+}t)-c_{-}\exp(-\frac{1}{2}
B_{-}t)
\label{eq:20}\, ,
\eeq
%-------------------------------------------------------------------------
%
%
where $c_{+}$ and $c_{-}$ are the contributions to the amplitude
from the region below and above the node position
with the corresponding effective diffraction slopes $B_{+}$ and $B_{-}$,
respectively ($B_{+} < B_{-}$).
Inequality $c_{+} > c_{-}$ means the 
undercompensation
whereas $c_{+} < c_{-}$ the overcompensation regime.
The destructive interference of these two contributions
results in a decrease of the effective diffraction slope
for $V'(2S)$
meson production towards small $t$ in contrary to the familiar increase
for the $V(1S)$ meson production.
Such a situation is shown in Fig.~7, where we present the model
predictions for the differential cross section
$d\sigma(\gamma^{*}\rightarrow V(V'))/dt$
for production of $V(1S)$ and $V'(2S)$ mesons at
different c.m.s. energies $W$ and at $Q^{2}=0$.

%
%
%------------------------- FIG.7. ----------------------------
%
%  
\begin{figure}[tbh]
  \includegraphics{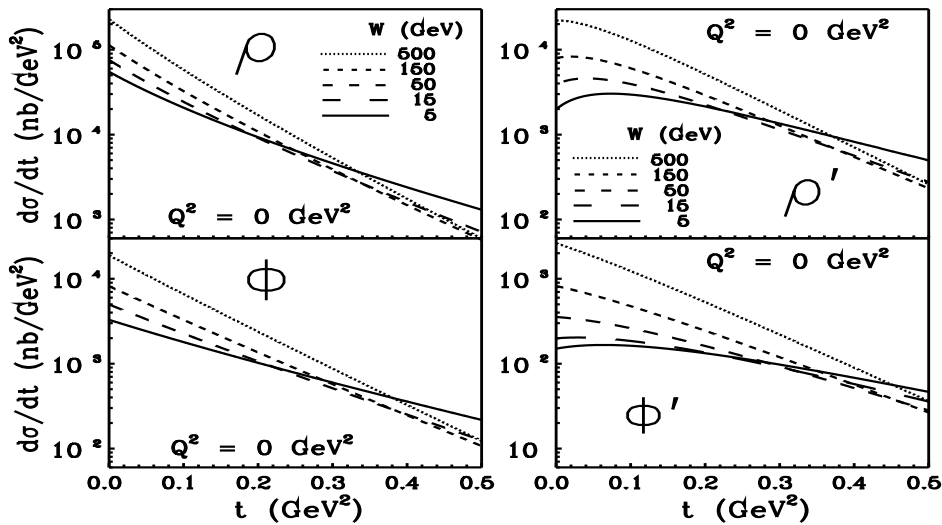}
  \begin{center}
  \vspace{6.3cm}
  \parbox{13cm}
  {\caption[Delta]
  {
~- The color dipole model
predictions for the differential cross sections
$d\sigma(\gamma^* \rightarrow V(V'))/dt$
for the real photoproduction ($Q^2=0$)
of the $\rho^{0}, \rho'(2S), \phi^{0}$ and $\phi'(2S)$
at different values of the c.m.s. energy
$W$.
}
%-------------------
       \label{event7}}
%-------------------
  \end{center}
  \end{figure}
%------------------------------------------------------------- 
%

Real photoproduction measures the purely transverse
cross section.
As was already mentioned,
$V'(2S)$ production amplitude is in undercompensation
regime (${\cal D}> 0$).
However, 
because of $r_{B} > r_{S}$,
the numerator ${\cal N}< 0$
at $W\lsim 150$\,GeV for $\rho'(2S)$ production
and at $W\lsim 30$\,GeV for $\phi'(2S)$ production.
As the result, the diffraction
slope is negative valued at $Q^{2}=0$.
At $t > 0$ the node effect becomes weaker.
The higher is $t$ the weaker is the node effect as a consequence
of the destructive interference (\ref{eq:20}) described above.
Consequently,
$d\sigma(\gamma^{*}\rightarrow V'(2S))/dt$ firstly rises
with $t$, flattens off at $t\in (0.0-0.2)$\,GeV$^{2}$
having a broad maximum.
At large $t$, the node effect is weak
and $d\sigma(\gamma^{*}\rightarrow V'(2S))/dt$
decreases with $t$ monotonously as
for $V(1S)$ production. 
The position of the maximum can be roughly evaluated
from (\ref{eq:20})
%
%
%-----------------------------------------------------------------------
\beq
t_{max} \sim \frac{1}{2(B_{-}-B_{+})}log\Biggl [\frac{c_{-}^{2}}{c_{+}^{2}}
\frac{B_{-}^{2}}{B_{+}^{2}}\Biggr ]
\, ,
\label{eq:21}
\eeq
%------------------------------------------------------------------------
%
%
with the supplementary condition
%
%
%------------------------------------------------------------------------
\beq
\frac{c_{-}}{c_{+}} > \frac{B_{+}}{B_{-}}
\label{eq:22}
\eeq
%------------------------------------------------------------------------
%
%
If the condition (\ref{eq:22}) is not fulfilled
$d\sigma(\gamma^{*}\rightarrow V'(2S))/dt$
has no maximum and exhibits a standard
monotonous $t$- behaviour. 
The predicted nonmonotonic $t$- behaviour of $d\sigma/dt$
for $\rho'(2S)$ and $\phi'(2S)$ production in the
photoproduction limit is strikingly different
especially at smaller energies from
the familiar decrease with $t$ of $d\sigma(\gamma\rightarrow
\rho^{0}(1S))/dt$ and $d\sigma(\gamma\rightarrow \phi^{0}(1S))/dt$
(see Fig.~7).

At larger energies, $W\gsim 150$\,GeV for the
$\rho'(2S)$ photoproduction and
$W\gsim 30$\,GeV for
$\phi'(2S)$ photoproduction, the node effect   
becomes weaker the diffraction slope is positive valued
(both ${\cal N}> 0$ and ${\cal D}> 0$)
and 
the nonmonotonic $t$ dependence of the differential cross
section is changed for the monotonic one but still the
effective diffraction slope decreases slightly towards small $t$
(see Fig.~7).

%
%
%------------------------- FIG.8. ----------------------------
%
%  
\begin{figure}[tbh]
  \includegraphics{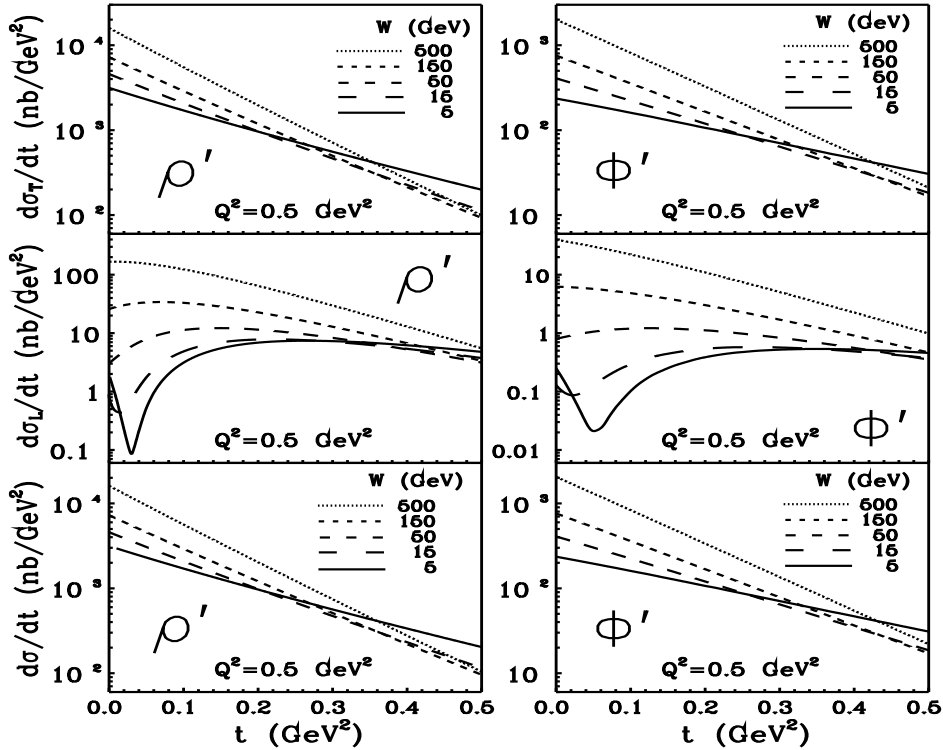}
  \begin{center}
  \vspace{9.5cm}
  \parbox{13cm}
  {\caption[Delta]
  {
~- The color dipole model
predictions for the differential cross sections
$d\sigma_{L,T}(\gamma^* \rightarrow V'(2S))/dt$ for
transversely (T)
(top boxes) and longitudinally (L)
(middle boxes) polarized radially excited
$\rho'(2S)$, $\phi'(2S)$
and for the polarization-unseparated
$d\sigma(\gamma^* \rightarrow V')/dt=
d\sigma_{T}(\gamma^* \rightarrow V'(2S))/dt+\epsilon
d\sigma_{L}(\gamma^* \rightarrow V'(2S))/dt$ for
$\epsilon = 1$ (bottom boxes)
at $Q^{2}=0.5$\,GeV$^{2}$ and different values of the c.m.s. energy
$W$.
}
%-------------------
       \label{event8}}
%-------------------
  \end{center}
  \end{figure}
%------------------------------------------------------------- 
%

Because of a possible overcompensation scenario
for $\rho'_{L}(2S)$ and
$\phi'_{L}(2S)$ mesons 
at small $Q^{2}$ (see also previous sections),
we present in Fig.~8 the model predictions for
$d\sigma(\gamma^{*}\rightarrow V'(2S))/dt$
at different energies $W$
and at fixed $Q^{2}= 0.5$\,GeV$^{2}$ for the production of
$(T)$, $(L)$ polarized and polarization unseparated $\rho'(2S)$ and
$\phi'(2S)$ mesons.
As it was mentioned above, at $Q^{2}= 0.5$\,GeV$^{2}$
the production amplitude for $\rho_{L}'(2S)$ and $\phi_{L}'(2S)$ mesons
is still in overcompensation regime ${\cal D}< 0$
(${\cal N}< 0$ as well) and the corresponding
diffraction slope $B(V_{L}'(2S))$ is positive valued at small energies 
$W\lsim 20$\,GeV.  
It results in a very spectacular pattern of anomalous $t$ dependence for
$d\sigma(\gamma^{*}\rightarrow V_{L}'(2S))/dt$
shown in Fig.~8 (middle boxes).
With rising $t$ due to above described destructive
interference
of two contributions to the production amplitude
(see (\ref{eq:20})),
one encounters the exact node effect
at some $t\sim t_{min}$.
Consequently, $d\sigma/dt$ firstly falls down
rapidly with $t$
having a minimum at $t\sim t_{min}$.
At still larger $t$
when the overcompensation scenario of $t$- dependent
production amplitude is changed for the undercompensation one
and the slope parameter becomes to be negative valued,
$d\sigma(\gamma^{*}\rightarrow V_{L}'(2S))/dt$ starts to rise with $t$
and further pattern of $t$- behaviour is analogical to that
for $V_{T}'(2S)$ production (see Fig.~7).

The position of the minimum $t_{min}$
is model dependent and can be roughly estimated from (\ref{eq:20})
%
%
%-------------------------------------------------------------------
\beq
t_{min} \sim \frac{1}{2(B_{-}-B_{+})}log\Biggl [\frac{c_{-}^{2}}{c_{+}^{2}}
\Biggr ]\, .
\label{eq:23}
\eeq
%------------------------------------------------------------------------
%
%

The gBFKL model predictions give
$t_{min}\sim 0.03$\,GeV$^{2}$ for $\rho_{L}'(2S)$ production
and
$t_{min}\sim 0.05$\,GeV$^{2}$ for $\phi_{L}'(2S)$ production
at $Q^{2} = 0.5$\,GeV$^{2}$ and at $W= 5$\,GeV.
However, we can not exclude a possibility that
this minimum will take a place at other values of $t$.   
At $Q^{2} < 0.5$\,GeV$^{2}$, $t_{min}$ will be
located at larger values of $t$.
At higher energy, the position of $t_{min}$ is shifted
to a smaller value of $t$ unless the exact
node effect is reached at $t=0$.
At still larger energy, when the
$V'_{L}(2S)$ production amplitude is in undercompensation regime,
this minimum disappears and
we predict the pattern of $t$- behaviour for
$d\sigma(\gamma^{*}\rightarrow V_{L}'(2S))/dt$
analogical to that for
$d\sigma(\gamma\rightarrow V_{T}'(2S))/dt$
in the photoproduction limit depicted in Fig.~7.
These predicted anomalies can be tested at HERA measuring
the diffractive electroproduction of $V'(2S)$
light vector mesons in separate polarizations $(T)$ and $(L)$.

%********************
\section{Conclusions}
%********************

We study the
diffractive photo- and electroproduction
of radially excited $V'(2S)$
vector mesons within the color dipole gBFKL dynamics.
We present a short review of possible anomalies,
which can be presently observed at HERA experiments.
The predicted anomalies are connected with the
node position in $V'(2S)$ radial wave function.

Firstly, we present a
rich pattern of anomalous $Q^{2}$ 
and energy dependence of the production cross section.
As a consequence of the node effect especially for
light vector mesons ($\rho'(2S), \phi'(2S)$)
we predict a very strong suppression of the
$V'(2S)/V(1S)$ production ratio in the real photoproduction limit of
very small $Q^{2}$.
For the longitudinally polarized $V'(2S)$
mesons we find a plausible overcompensation scenario leading to a sharp
dip of the longitudinal cross section
$\sigma_{L}(2S)$ at some finite $Q^{2} 
=Q_{c}^{2}\sim 0.5$\,GeV$^{2}$. The position $Q_{c}^{2}$ of this dip
depends on the energy and leads to a nonmonotonic energy
dependence of $\sigma_{L}(2S)$ at fixed $Q^{2}$.
At larger $Q^{2}$ and smaller scanning radius
we predict a steep rise of the $V'(2S)/V(1S)$ cross section ratio.
The flattening of this $2S/1S$ ratio at large $Q^{2}$ is
a non-negotiable prediction from the color dipole dynamics.

Further predictions are related to the
anomalous pattern of $Q^{2}$ and energy behaviour of the
diffraction slope for the production of $V'(2S)$
vector mesons. 
At moderate energies,
we find a nonmonotonic $Q^{2}$ dependence of the slope
parameter which can be tested at HERA in the range  
of $Q^{2}\in (0-10)$\,GeV$^{2}$.
For the production of $(L)$ polarized $V'(2S)$
as a consequence of the 
overcompensation scenario we find a sharp peak followed
immediately by
a very rapid transition of the slope parameter   
from positive to negative values
at $Q^{2}\sim Q'^{2}_{L}\in (0.5-1.5)$\,GeV$^{2}$
The position of this rapid transition $Q'^{2}_{L}$
is energy dependent and leads
to a nonmonotonic energy dependence of $B(V_{L}'(2S))$ at fixed
$Q^{2}$.
At $Q^{2}=0$ when
the node effect is strong, for the undercompensation
scenario we predict a counterintuitive
inequality $B(V'(2S)) < B(V(1S))$.
However, for overcompensation
scenario we predict the expected standard inequality
$B(V'(2S)) > B(V(1S))$.
This is a very crucial point of a possible experimental
determination of a concrete scenario
extracting from the data at HERA the $B(V'(2S))$ diffraction slope
in the photoproduction limit.

The last class of predictions concerns to an
anomalous $t$ dependence of the $V'(2S)$ 
differential cross section.
The origin is in
destructive interference of the
large distance negative contribution to the
production amplitude from the region
above the node position with a steeper
$t$- dependence and the  
small distance positive contribution to the
production amplitude from the region
below the node position with a weaker
$t$- dependence.
As the result,
we predict at $Q^{2}=0$
a nonmonotonic $t$ dependence of
$d\sigma(\gamma\rightarrow V_{T}'(2S))/dt$
and a decreasing diffraction slope for $V_{T}'(2S)$ mesons
towards small values of $t$
in contrary with the familiar increase of $B(V(1S))$
for $V(1S)$ vector mesons.
The differential cross section
firstly rises
with $t$ having a broad maximum.
At larger $t$ when the node effect is still
weaker, $d\sigma(\gamma\rightarrow V_{T}'(2S))/dt$ has the standard
monotonic $t$- behaviour as for production of $V(1S)$
vector mesons.
For production of $(L)$ polarized $V_{L}'(2S)$ mesons,
there is overcompensation at $t=0$ leading to
a dip (minimum) of the differential cross section at $t\sim t_{min}$.
The position of $t_{min}$ is
energy dependent and is the result of the model dependent
soft-hard cancellations.

To conclude, if the data will exhibit 
a dip (minimum) in energy ($Q^{2}$) dependence of the 
$V'(2S)$ production cross section, $V'(2S)/V(1S)$ cross
section ratio, diffraction slope and in $t$ dependence
of the differential cross section $d\sigma/dt$ then
the corresponding $V'(2S)$ production amplitude
is in the overcompensation regime.
Otherwise the $V'(2S)$ production amplitude is
in the undercompensation regime.

%
%
% --------------------------- REFERENCES -------------------------------
%
%  

%
%
%-----------------------------------------------------------------------
%
%
  
\end{document}